\PassOptionsToPackage{table}{xcolor}
\documentclass[sigconf]{acmart}
\usepackage{booktabs} 
\usepackage[singlelinecheck=false]{caption}

\let\oldFootnote\footnote
\newcommand\nextToken\relax

\renewcommand\footnote[1]{%
    \oldFootnote{#1}\futurelet\nextToken\isFootnote}

\newcommand\isFootnote{%
    \ifx\footnote\nextToken\textsuperscript{,}\fi}

\setcopyright{rightsretained}

\acmConference[PROMISE'19]{The 15th International Conference on Predictive Models and Data Analytics in Software Engineering}{September 2019}{Brazil}
\acmYear{2019}
\copyrightyear{2019}

\begin{document}
\title[NPM User Participation and User Classification]{Patterns of Effort Contribution and Demand and User Classification based on Participation Patterns in NPM Ecosystem}

\author{Tapajit Dey}
\orcid{0000-0002-1379-8539}
\affiliation{%
  \institution{University of Tennessee, Knoxville}
  \city{Knoxville}
  \state{Tennessee}
}
\email{tdey2@vols.utk.edu}

\author{Yuxing Ma}
\affiliation{%
  \institution{University of Tennessee, Knoxville}
  \city{Knoxville}
  \state{Tennessee}
}
\email{yma28@vols.utk.edu}

\author{Audris Mockus}
\affiliation{%
  \institution{University of Tennessee, Knoxville}
  \city{Knoxville}
  \state{Tennessee}
}
\email{audris@utk.edu}


\begin{abstract}
    Background: Open source requires participation of volunteer and commercial 
    developers (users) in order to deliver functional high-quality components.
    Developers both contribute effort in the form of patches and demand effort from the component maintainers to resolve issues reported against it.
    Aim: Identify and characterize patterns of effort contribution and demand throughout the open source supply chain and investigate if and how these patterns vary with developer activity; identify different groups of developers; and predict developers' company affiliation based on their participation patterns.
    Method: 1,376,946 issues and pull-requests created for 4433 NPM packages with over 10,000 monthly downloads and full (public) commit activity data of the 272,142 issue creators is obtained and analyzed and dependencies on NPM packages are identified. Fuzzy c-means clustering algorithm is used to find the groups among the users based on their effort contribution and demand patterns, and Random Forest is used as the predictive modeling technique to identify their company affiliations.
    Result: Users contribute and demand effort primarily from packages that they depend on directly with only a tiny fraction of contributions and demand going to transitive dependencies. A significant portion of demand goes into packages outside the users' respective supply chains (constructed based on publicly visible version control data). Three and two different groups of users are observed based on the effort demand and effort contribution patterns respectively. The Random Forest model used for identifying the company affiliation of the users gives a AUC-ROC value of 0.68. 
    Conclusion: Our results give new insights into effort demand and supply at different parts of the supply chain of the NPM ecosystem and its users and suggests the need to increase visibility further upstream.
\end{abstract}

%
%



\maketitle

\section{introduction}\label{s:intro}
Open Source Software is characterized by the fact that the source code is publicly 
available and can be modified and reused with limited restrictions by the public.
This has led to the creation of user communities that contribute regularly to the 
development process~\cite{lakhani2004open,von2001learning}, primarily through 
reporting and, in many cases, fixing bugs~\cite{ducheneaut2005socialization}. 
Reported bugs, in effect, create a demand for effort needed to address them, and, 
it has been extensively documented (see, e.g.~\cite{xie2013impact}) that large numbers 
of low-quality issues may overwhelm the projects. Some users also provide patches with 
their issues (pull requests or PRs) which should require, at least in principle, much 
less effort to address, and can be regarded as a contribution of effort to the project 
(the effort spent by the user to create the patch). We refer to PRs when talking about effort
contribution and issues without patches when talking about the demand of effort in the 
further discussion. Since our data is collected from GitHub, which treats pull requests 
as issues, we follow the same terminology in our paper, i.e. when we talk about ``issues'', 
we refer to both issues with and without patches. Many of the contributors, who create these issues and patches, are potential developers and/or developers of their own projects. When we refer to ``users'' in this paper, we actually are referring to this population of user-developers.

Software ecosystems, by their nature, enable creativity and productivity by letting 
developers not to write software from scratch but only focus on incremental improvements 
that depend on other modules in the ecosystem for bulk of the functionality. This 
results in a complex supply chain of dependencies within the ecosystem. The supply chain of 
a user consists of the direct as well as transitive dependencies on repositories to which they maintain. While some 
studies found that contributing and demanding effort is more complicated when 
it crosses project boundaries in direct and transitive
dependencies~\cite{bogart2016break,valiev2018ecosystem}, it is not clear how 
prevalent such contribution and demand is at the ecosystem level.  

This question 
is closely associated with the concept of visibility~\cite{amreen2019methodology} within 
a supply chain, which refers to how far a user can ``see'' in the supply chain beyond 
their direct upstream dependencies, i.e. if they are aware of the transitive dependencies 
of the projects they are using. This is an important question since a lack of visibility 
in an ecosystem is detrimental to the users' capacity to contribute, leading to a limitation
in user innovation,  potential licensing conflicts due to a transitive dependency using a different 
license, exposing users to higher risk due to the user not being aware of bugs
upstream that can be used to instigate a supply chain
attack~\footnote{https://it.slashdot.org/story/19/06/08/1940204/how-npm-stopped-a-malicious-upstream-code-update-from-stealing-cryptocurrency}\footnote{https://www.bleepingcomputer.com/news/security/somebody-tried-to-hide-a-backdoor-in-a-popular-javascript-npm-package/} and various 
other types of risks (see, e.g.~\cite{boehm1991software,wallace2004understanding,rigby2016quantifying,alberts2011systemic}), and various other problems. Visibility within a supply chain is not easy to measure, however, the number of cross-project issues and PRs is a good proxy. An ecosystem with greater visibility would allow its users to be able to contribute to their transitive dependencies more frequently.  Therefore, by measuring where the issues and PRs are concentrated in the supply chain, we can get a good sense about the level of visibility within the ecosystem.
Thus, we present our first research question as:\\
\textbf{RQ1: Where in the supply chain do the contribution of effort and demands for effort
 occur?}

We are also interested in discovering if we can identify different groups of users based on their participation patterns, i.e. in which layer of their respective supply chains they contribute effort or demand effort from. 
The answer to this question is important to identify
and characterize different sub-communities of users within the ecosystem. So, the second research question we are addressing in this paper is: \\
\textbf{RQ2: Can we identify different groups among the users based on their participation patterns?}

It has been previously observed that the so called one-time-contributors~\cite{lee2017understanding,lee2017one} might have different motivation and behave differently from more involved participants. We, therefore, would like to understand if such distinctions apply in large software ecosystems, i.e. if more active developers contribute different proportion of their effort to upstream projects than casual users. We identify the more prolific users as those who have submitted at least 10 issues to the ecosystem under consideration (we are not counting the issues submitted to other ecosystems). The number 10 is somewhat arbitrary, but, given that 75\% of the users in our sample submit 3 or fewer issues, it only includes individuals representing less than ten percent of all users. \\
\textbf{RQ3: Do the answers of RQ1 and RQ2 change if we consider only the more prolific users?}

Finally, commercial entities tend to participate in FLOSS in ways that are distinct from the way volunteer or independent developers participate~\cite{zhou2016inflow}, which stem from a number of facts, like differences in motivation, interest, urgency, and expertise.  Therefore, we would like to understand if such distinctions apply in large software ecosystems, and, in turn, if the participation patterns can be used to predict if a user has a commercial affiliation. GitHub user profiles have the option of declaring if a user works for a company, and, it can be argued that more serious users take their time to populate their profiles accurately. However, the Git version control system extends far beyond GitHub, and a model that can  identify the commercial affiliation of a user by looking into their participation (issue and PR creation) patterns would be useful in classifying the types of users in platforms that do not have this option. Thus, our last research question is:\\
\textbf{RQ4: Can we use the participation patterns of users to predict their commercial affiliation?}

We chose node package manager (NPM) to answer our research questions because of the size of the ecosystem, availability of data, and the large number of its users who work for a company. NPM is a package manager of JavaScript packages, and is one of the largest OSS communities at present, with over 932,000 different packages (Apr, 2019) and millions of users (estimated 4 million in 2016~\cite{npmuser}, and about 4000 new users on an average day\footnote{https://twitter.com/seldo/status/880271676675547136}. NPM is used heavily by companies. According to the NPM website\footnote{https://www.npmjs.com/}, all 500 of the Fortune 500 companies use NPM, and they claim that: `` \textit{Every company with a website uses npm, from small development shops to the largest enterprises in the world.}'' Given the heavy industry use of NPM, a good number of the users who contribute to it are likely to have a commercial affiliation, which should give us a more balanced dataset to answer \textbf{RQ4}. However, most packages in NPM are not widely used and have limited or no issues or PRs. We, therefore, focused on 4433 NPM packages with over 10,000 monthly downloads since January, 2018, that also had an active GitHub repository with at least 1 issue. All issues ever filed against these packages were obtained using the GitHub API (pull-requests are treated as issues by the GitHub API) resulting in 1,376,946 issues and PRs, out of which 541,715 (39\%) were pull-requests. We also retrieved information for the 272,142 still active users (some users who filed issues had deleted their accounts and had their id replaced the special GitHub id ``ghost'').  

Our primary findings are: (1) Users are more likely to contribute issues and PRs to their direct dependencies, but a number of issues were created for packages outside a user's supply chain, and very few cross-project issues and PRs were observed. (2) Three different user groups were observed based on the users' effort demand patterns, those who are likely to create issues to their direct dependencies, those who are likely to create issues to packages none of their public repositories depend on, and a small group of users who are likely to create cross-project issues. Based on the effort contribution patterns we observed two major groups, similar to the first two groups observed based on the effort demand pattern. (3) We see that more prolific users are even more likely to contribute to their direct dependencies and much less likely to contribute to packages outside their respective supply chains. (4) We were able to identify the company affiliations of the users with 70\% accuracy (95\% Confidence Interval between 69.9\% and 70.54\%) with their contribution patterns as predictors using a tuned Random Forest model, with the value of AUC under the ROC curve being 0.68.

The rest of the paper is organized as follows: In Section~\ref{s:relwork}, we discuss the related works in the topic. In Section~\ref{s:method}, we discuss the Methodology, focusing on the analysis method we followed. In Section~\ref{s:data}, we describe the data collection  and data processing steps, focusing on the design choices made along the way. In Section~\ref{s:result}, we describe the results we found pertaining to our research questions. The implications of the findings is discussed in Section~\ref{s:disc}. Finally, we discuss the limitations of our study in Section~\ref{s:limit} and conclude our paper in Section~\ref{s:conclusion}.

\vspace{-10pt}
\section{Related Work}\label{s:relwork}

The NPM ecosystem is one of the most active and dynamic JavaScript ecosystems and~\cite{wittern2016look} presents its dependency structure and package popularity. Studies on NPM have mostly focused on its dependency networks~\cite{decan2018impact}, its effect on popularity of NPM packages~\cite{dey2018software}, and problems associated with library migration~\cite{zapata2018towards}.

As a part of our study we look at the dependencies of the JavaScript projects in GitHub, and the different NPM packages. However, we look not only at the direct dependencies, but also into the transitive dependencies of the packages, i.e. dependencies of dependencies of the packages. A number of studies looked into the handling of dependencies of NPM ecosystem in particular. E.g.,~\cite{zerouali2018empirical} conduct an empirical study on the lag in updating a package in conjunction to its dependencies in NPM and its effect, while~\cite{decan2017empirical} conduct an comparative study of dependency handling by NPM, R-CRAN, and RubyGems ecosystems, and compare the different strategies used by the three in handling dependency updates. 

Our first research question looked into the aspect of issue reporting and the prevalence of cross-project issues in NPM ecosystem. The number of observed issues and PRs is directly dependent on the amount of usage, as reported in~\cite{dey2018modeling}. ~\cite{decan2016github} showed that failures in upstream packages brought more and more troubles to the downstream projects. An approach to identify Cross-System-Bug-Fixings in FreeBSD and OpenBSD kernels was proposed by~\cite{canfora2011social}. Other studies in this topic explored how the downstream developers find the root causes and coordinate with upstream developers to fix the problems~\cite{ma2017developers}, the workarounds employed by downstream developers when faced with a bug in an upstream project~\cite{ding2017empirical}, and the question of how to automate the fix of a bug introduced by a third party library upgrade~\cite{ding2017empirical}. Unlike these studies, we focus on both effort demand and supply and employ a much larger data-set of projects.

One of our research questions center around predicting users with a company affiliation based on  the differences in the types of contributions. Just because a user is affiliated to a company doesn't necessarily imply that they use the NPM packages for their job applications, but it may increase that likelihood. Our belief in this assumption is bolstered by the result of the 2018 Node.js User Survey Report\footnote{https://nodejs.org/en/user-survey-report/}, which found that: \textit{``A majority} (of users of NPM packages) \textit{are developers (as opposed to dev managers), in small ($<$100 employees) companies, with 5+ years of professional development experience.''} Given the typical user base, we believe it is a fair assumption that a significant number of users who have disclosed that they have a company affiliation, actually use these packages as a part of their day job and not as a hobby. 

FLOSS development started with the goal of emphasizing the freedom of computer users\footnote{https://www.gnu.org/philosophy/floss-and-foss.en.html}. Although initially the commercial software development community steered clear of open source software, its benefits, as discussed in studies like~\cite{mockus2002two}, soon led them into using and supporting open source software development. A plethora of studies looked into the scenario of commercial adoption of open source software, e.g.~\cite{glynn2005commercial,chau1997factors}
to name a few. Currently, the interaction between open source software and different software companies is much stronger and closer, with many companies actively supporting open source development, and using different open source software in a daily basis. Although a number of studies looked into the benefits of using open source software by a company(e.g.~\cite{von2001learning}), and the result of commercial involvement~\cite{ciesielska2016dilemmas,zhou2016inflow}
by studying different project level metrics like sustainability, developer inflow and retention etc., to our knowledge no study has looked into the difference in types of contribution of individual commercial and non-commercial users on a large scale software ecosystem like NPM and used it for predicting if a user has commercial affiliation.

\vspace{-10pt}
\section{Methodology}\label{s:method}
In this section, we discuss some terminologies we used in this study and discuss the analysis method we followed.

\vspace{-10pt}
\subsection{Terminologies}\label{ss:term}
Our research questions look into the packages where a user creates issues and PRs, and at which level of the user's supply chain these packages belong to, and we define some terminologies describing these levels for the ease of referring to these levels. 

The NPM packages that a user(developer) contributes to directly are referred as \textit{level 0} packages for that user, i.e. only users who have committed to an NPM package directly, and not through a pull request, can have \textit{level 0} packages. Arguably, these user-developers are part of the core team of that NPM package, since they have direct write access to that repository.

The direct dependencies of all repositories a user has ever committed to (we utilize a recent version of WoC data~\cite{woc19} to collect information from all repositories, including projects that are not registered in NPM) are called \textit{level 1} packages for that user. Furthermore, \textit{level 1} packages also includes originating packages that the said user has forked. 

The direct and transitive dependencies of the \textit{level 1} packages are classified as \textit{level 2+} packages of the user. Contributions to \textit{level 2+} packages can be regarded as cross-project contributions by the said user, since these are transitive dependencies for them. The reason we referred to level 2 or higher packages by aggregating them into \textit{level 2+} is that the number of reported issues dropped drastically starting from level 2. Moreover, since any issue reported at level 2 onward would be qualified as a cross-project issue, such aggregation seemed reasonable. 

The remaining packages in NPM ecosystem are \textit{level X} packages for that user, since these include all the packages none of the public repositories the user has committed to depend on even transitively. For obvious reasons, we could only observe the publicly visible repositories the user-developer committed to. These packages are the ones that are outside a user's supply chain, but for the sake of consistency and ease of referring, we call them \textit{level X} packages.

The issues and PRs created by a user for a package which belongs to one of these levels of the supply chain for that user are regarded as the issues and PRs created for that level by that user.

\vspace{-10pt}
\subsection{Analysis Method}\label{ss:an}
The data collection was done using Python, and the analysis was performed using R. 

We started by collecting the necessary data, which was used to create our final dataset. The data collection and data processing steps are described in detail in Section~\ref{s:data}.

Python scripts were used to create the data files necessary for analysis. We carefully tabulated the number of issues and PRs created for each level of the supply chains of the users to address our first research questions.

To answer RQ2, we decided to calculate the marginal probabilities of each user creating an issue and a PR to each level in their respective supply chains. 
However, we observed only around 1 in 3 users create a PR, and looking into the two probabilities together would have automatically put 2/3rds of the users in one group and the rest in other. So, we decided to look only at the probabilities of users creating issues (at different levels in their respective supply chains) when looking at all users, and look at the probabilities of users creating PRs (at different levels in their respective supply chains) only for the subset of users who have created at least one pull request.

We used the fuzzy c-means clustering algorithm~\cite{bezdek1984fcm} for answering RQ2. We decided to use this instead of the more commonly used k-means or hierarchical clustering algorithm because we suspected, and later observed, that there is a lot of overlap in our data, and k-means doesn't work well with such data; as for hierarchical clustering, given we have 272,142 users in our dataset, calculating the distance matrix needed to construct the clusters proved very difficult due to the computational resources required.  The fuzzy c-means algorithm assigns membership probabilities to each data point instead of assigning them to clusters directly, which gives the best results for the type of data we have. We used the fuzzy c-means implementation in the \textit{e1071} R package, and for visualizing the clusters we used the ``clusplot'' function in the \textit{cluster} R package.

We used Random Forest model (\textit{randomForest} package) for training our predictive model (RQ4), since it is one of the best performing models. The model parameters (``ntree" and ``mtry") were tuned using functions from \textit{caret} and \textit{e1071} packages.

\vspace{-10pt}
\section{Data Description}\label{s:data}
In this section, we describe the data collection  and data processing steps, focusing on the design choices that were made along the way.

\vspace{-10pt}
\subsection{Data Collection}
Keeping our research questions in mind, we needed the following types of data:  
\begin{enumerate}
    \item The list of NPM packages that satisfy our criteria of having more than 10,000 downloads per month and a GitHub repository with at least one issue.
    \item Link to GitHub repositories of these packages for collecting the issues. 
    \item List of all issues and issue creators of these packages.
    \item Detailed information on the issue creators to know if they disclose their company affiliation.
    \item List of all commits made by these users, and the list of GitHub repositories where they made those commits.
    \item List of source repositories of the forked repositories the users may have committed to.
    \item List of all dependencies (NPM packages) of the GitHub repositories the users committed to.
    \item List of dependencies of all NPM packages for creating the transitive list of dependencies for the repositories the users committed to.
\end{enumerate}
\vspace{-10pt}

The data for item (1) was collected from the \texttt{npms.io} website, using the API provided~\footnote{https://api.npms.io/v2/package/[package-name]}. The associated GitHub repository URL (item 2) and the list of dependencies of the NPM packages (used for item 8) were collected from their metadata information, which was obtained by using a ``follower" script, as described in NPM's GitHub repository~\footnote{https://github.com/npm/registry/blob/master/docs/follower.md}. After filtering for our criteria that the NPM package must have more than 10,000 monthly downloads (since January, 2018), a functional link to its GitHub repository, and at least one issue, we were left with 4433 different NPM packages. 

The list of all issues for the packages (item 3) was obtained using the GitHub API for issues\footnote{https://developer.github.com/v3/issues/}, using the \texttt{state=all} flag. We ended up with 1,376,946 issues (until January, 2019, when the data was collected) for the 4433 packages. It is worth mentioning here that sometimes more than one NPM package can have the same associated GitHub repository, e.g. all TypeScript NPM packages (starting with ``@types/'', like @types/jasmine, @types/q, @types/selenium-webdriver etc.) refer to GitHub repository\\ ``DefinitelyTyped/DefinitelyTyped''. To avoid double-counting and further confusion, we saved the issues keying on the repository instead of the package name, though we also saved the list of packages associated with a repository. We found that there are 3797 unique repositories associated with these 4433 packages. 

Then we extracted the list of all users who created these issues and obtained detailed information on them (item 4) using the GitHub API\footnote{https://developer.github.com/v3/users/}. We found that there were 272,142 users still active (as of March, 2019, when the data was collected) out of 280,835 users who had created issues for the NPM packages under consideration. 

For obtaining information on items (5) and (6), we used the GHTorrent database~\cite{Ghtorrent} available in the Google Cloud platform\footnote{http://ghtorrent.org/gcloud.html} (we used the \texttt{ghtorrent-bq:ght\_2018\_04\_01} database), and extracted the relevant information using Google BigQuery. 

To get a list of all projects a user ever committed to (item 5), we extracted the list of commits made by a user and got the list of the repositories where those commits were made, finally getting the list of all repositories  the user committed to. We found that the 272,142 users committed to 6,676,089 projects in total, and it had a very skewed distribution in terms of the number of projects a user committed to. Note that these projects don't have to be JavaScript projects, since we obtained this information from all Git data~\cite{woc19}. Upon further analysis, it was found that 5,898,782 of them had a \texttt{package.json} file, so we classified them as JavaScript projects, and used them for further analysis. 

For getting the sources of the forked repositories the users might have committed to (item 6), we used the \textit{projects} table in the GHTorrent database, which has a field named ``forked\_from'', and performed a recursive search (since project A can be forked from B, and B can be forked from C etc.) to get the list of all sources.

For the data in item (7), we extracted information for all GitHub repositories that has a \texttt{package.json} file and extracted the dependency information from that. We also found that some repositories use another file named \texttt{lerna.json} to list their dependencies. So, we extracted dependency information from this file as well where it was available. 

There were cases where the users directly committed to a package repository. Those were treated as special cases and handled using a map of package name and package URL  constructed previously. 

The transitive dependency map of item (8) was constructed by doing a recursive search using the dependency information collected for the packages. We listed the direct dependencies of a package as level 1 dependencies of that package, the dependencies of the packages in level 1 as level 2 dependencies of that package, and so on. It is worth mentioning that if a package A, for example, was found to be dependent on a package B directly, as well as through another package C ( A depends on C, C depends on B), we took the lower number, i.e. B was still listed as level 1 dependency of A. Moreover, although forks are not dependencies of a project in the same way other dependencies work, we decided to add the sources of the forked repositories as level 1 dependencies for ease of representation. However, from level 2 onward, we only have packages in the list of dependencies, which includes the dependencies of the source repositories of the forked ones. 

\vspace{-10pt}
\subsection{Data Processing}

\begin{table*}[htb]
\caption{Final List of Variables in the Dataset}\label{t:var}
\vspace{-10pt}
\resizebox{\textwidth}{!}{%
{\footnotesize 
\begin{tabular}{|p{4cm}| p{4cm}| p{4cm}| p{4cm}|}
\hline
User login   \$                                               & No. of projects the user committed to                                 & No. of repos that are forks of other repos                  & No. of NPM packages committed to                            \\\hline
No. of direct dependencies of all the user's packages       & No. of transitive dependencies of all the user's packages    & Total no. of issues created by the user                     & Total no. of PRs created by the user                        \\\hline
No. of issues created for level 0 packages                  & No. of issues created for level 1 packages                   & No. of issues created for level 2+ packages                 & No. of issues created for level X packages                  \\\hline
No. of PRs created for level 0 packages                     & No. of PRs created for level 1 packages                      & No. of PRs created for level 2+ packages                    & No. of PRs created for level X packages                     \\\hline
Total no. of packages for which a issue was created         & Total no. of packages for which a PR was created             & No. of level 0 packages for which an issue was created      & No. of level 0 packages for which a PR was created          \\\hline
No. of level 1 packages for which an issue was created      & No. of level 1 packages for which a PR was created           & No. of level 2+ packages for which an issue was created     & No. of level 2+ packages for which a PR was created         \\\hline
No. of level X packages for which an issue was created      & No. of level X packages for which a PR was created           & Total no. of issues that are not pull requests              & No. of non-pull-request issues created for level 0 packages \\\hline
No. of non-pull-request issues created for level 1 packages & No. of non-pull-request issues created for level 2+ packages & No. of non-pull-request issues created for level X packages & If the user has a company affiliation  \$      \\\hline
\end{tabular}%
}}
\vspace{-10pt}
\end{table*}

The raw data was processed to create a usable dataset for analysis. For each user, we first extracted the list of repositories they contributed to and then constructed the list of packages they transitively depend on. The transitive (level 2+) dependencies for a user was calculated using the transitive list of dependency data (item (8) above). Then we extracted the packages the user had raised issues for, and observed if that package belongs to level 0, 1, 2+, or X for that user. 

We noticed that the user id that created issues to the most number of packages was found to be ``ghost'', which is of little surprise, and it was removed from subsequent analysis. The second and third positions were occupied by two bots associated with the automated dependency management website/service Greenkeeper\footnote{https://greenkeeper.io/}, both of which raised issues for more than 400 different packages, and created pull-requests for 98\%  of those packages, and 92\% of the issues raised by these two bots were pull-requests. We further noticed that bots tend to create a lot more issues and PRs compared to human users. So,  we decided to remove the users that we could identify as bots, because bots are much more prolific by design, and could skew the distributions significantly. We were able to identify 35 bots which were removed from further analysis. 

The variables in our final dataset are listed in Table~\ref{t:var}. Each entry in the table is the observation for one user. All variables, except User login and whether the user has company affiliation (marked by \$ in Table~\ref{t:var}), are numerical in nature.

\vspace{-10pt}
\section{Results}\label{s:result}

\begin{table*}[htb]
\caption{Distribution of Issues created by Users for different levels in their respective supply chains \newline Numbers on the right show the values for users with 10 or more issues, pertaining to RQ3}
\label{t:i}
\vspace{-10pt}
\resizebox{\textwidth}{!}{%
\begin{tabular}{p{4.5cm}| p{3cm}| p{3cm}| p{3.25cm}| p{3.25cm}}
\hline\hline
 & Fraction of issues created for Level 0 & Fraction of issues created for Level 1 & Fraction of issues created for Level 2+ & Fraction of issues created for Level X \\ \hline\hline
All users who created an issue & \cellcolor{blue!20}0.027 | 0.039 & \cellcolor{blue!20}\textbf{0.532 | 0.688} & \cellcolor{blue!20}0.039 | 0.039 & \cellcolor{blue!20}0.402 | 0.234 \\ \hline\hline
Users who created  issue for level 0 & 0.139 | 0.127 & \textbf{0.761 | 0.778 }& 0.028 | 0.028 & 0.071 | 0.067 \\ \hline
Users who created  issue for level 1 & 0.033 | 0.041 & \textbf{0.760 | 0.772} & 0.039 | 0.039 & 0.168 | 0.148 \\ \hline
Users who created  issue for level 2+ & 0.031 | 0.034 & \textbf{0.679 | 0.728} & 0.116 | 0.077 & 0.174 | 0.160 \\ \hline
Users who created  issue for level X & 0.019 | 0.029 & \textbf{0.456 | 0.652} & 0.035 | 0.042 & 0.490 | 0.278 \\ \hline
\end{tabular}%
}
\vspace{-10pt}
\end{table*}

\begin{table*}[htb]
\caption{Distribution of Pull Requests (PRs) created by Users for different levels in their respective supply chains \newline Numbers on the right show the values for users with 10 or more issues, pertaining to RQ3}
\label{t:pr}
\vspace{-10pt}
\resizebox{\textwidth}{!}{%
\begin{tabular}{p{4.5cm}| p{3cm}| p{3cm}| p{3.25cm}| p{3.25cm}}
\hline\hline
 & Fraction of PRs created for Level 0 & Fraction of PRs created for Level 1 & Fraction of PRs created for Level 2+ & Fraction of PRs created for Level X \\ \hline\hline
All users who created a PR & \cellcolor{blue!20}0.048 | 0.056 & \cellcolor{blue!20}\textbf{0.772 | 0.810} & \cellcolor{blue!20}0.020 | 0.015 & \cellcolor{blue!20}0.160 | 0.119 \\ \hline\hline
Users who created PR for level 0 & 0.171 | 0.155 & \textbf{0.791 | 0.809} & 0.009 | 0.009 & 0.029 | 0.027 \\ \hline
Users who created PR for level 1 & 0.047 | 0.057 & \textbf{0.884 | 0.881} & 0.014 | 0.014 & 0.054 | 0.049 \\ \hline
Users who created PR for level 2+ & 0.042 | 0.044 & \textbf{0.843 | 0.868} & 0.055 | 0.033 & 0.06 | 0.055 \\ \hline
Users who created PR for level X & 0.034 | 0.038 & \textbf{0.727 | 0.794} & 0.018 | 0.016 & 0.222 | 0.152 \\ \hline

\end{tabular}%
}
\vspace{-10pt}
\end{table*}

In this section we discuss our findings and answer the different Research Questions we had, staring with some general statistics about the data. Since our RQ3 is asking the same questions as our RQ1 and RQ2, but with a different condition, we present the answer of RQ3 together with the answers of RQ1 and RQ2.

\vspace{-15pt}
\subsection{General Statistics about the Data}\label{ss:gen}
Here we discuss some general statistics, which, in spite of not being directly related to our research questions, can give us some insight into the data and the NPM ecosystem in general.

To recap, our study focused on 4433 NPM packages (3797 unique GitHub repositories) with more than 10,000 monthly downloads since January, 2018. We collected 1,376,946 issues created for these projects, including 541,715 pull-requests, which were created by 280,835 users, out of whom 272,142 were active at time of data collection.

A few interesting statistics about the data are reported below:
\vspace{-10pt}
\begin{itemize}
    \item We found that 219,945, or around 81\% of the total users had committed to at least one public repository in GitHub.
    \item 84,813 (31\%) users have a disclosed company affiliation, but they created almost 57\% of the pull-requests, and around 42\% of the issues.
    \item 87,653 (32\%) users had created at least one pull request, or, 68\% of the users have created issues but never submitted a pull request.
    \item 38,080 (14\%) users have never submitted any issue without a patch, i.e. all the issues they submitted were PRs.
    \item 4585 (1.7\%) users in our user base had committed to at least one NPM package directly, so were likely part of the core team of an NPM package.
    \item 139,917 (51\%) users committed only issue, i.e. just over half of the users who committed at least one issue  were ``one-time-contributors'', and they create around 11\% of the total no. of issues, and around 4.6\% of the total no. of PRs.
    \item 215,584 (79\%) users committed at least one issue, and 31,330 (12\%) of the users committed at least one PR to a package not in their supply chain (level X).
    \item 21,144 (8\%) and 4643 (1.7\%) users committed at least one issue and at least one PR respectively, to a transitive dependency package. i.e. they submitted cross-project issues and cross-project pull requests respectively.
    \item 89,149 (33\%) and 62,262 (23\%) users committed at least one issue and at least one PR respectively, to a direct dependency package.
    \item Only 19,376 users had created more than 10 issues (corresponding to our condition in RQ3), which consists of roughly 7\% of the entire user population, but they create around 60\% of the total issues and 75\% of the total PRs. 
    \item All of the numerical variables listed in Table~\ref{t:var} have extremely skewed distribution.
\end{itemize}
\vspace{-10pt}

Previous studies of contribution patterns reported a layered structure of a core team, bug fixers, and bug reporters for individual projects (see, e.g.~\cite{crowston2003social,wagstrom2012roles}). We see a similar distribution of the users, with 1.7\% of the users likely to be part of the core team of some package, 32\%, who provide patches, could be thought of as bug fixers, and the rest, which consists of the majority of the user base, are issue reporters. This shows the premise of the onion model is valid at the ecosystem level as well.

\vspace{-10pt}
\subsection{{\normalsize Where in the supply chain are the contribution of effort and demands for effort concentrated? (RQ1) \texorpdfstring{\newline}{} Does the distribution change for the more prolific users? (RQ3)}}

To answer this question we looked at the number of issues and PRs created by each user at different levels of their supply chain, as defined in Section~\ref{ss:term}. The results of the finding are reported in Tables~\ref{t:i} and~\ref{t:pr}, where the distribution of issues and PRs created by users for different levels in their respective supply chains are reported in terms of the fraction of issues and PRs reported at each level. 
The values on the left side are the fractions for all users under consideration, and the values on the right side are the fractions for the more prolific users.

We observe from Table~\ref{t:i} that, when considering all users, most of the issues (53.2\%) are reported for the direct dependencies of the users, followed by issues created (40.2\%) for packages on which none of the users' public repositories depend on. The fraction of cross-project issues is pretty small (3.9\%), and so is the number of issues created for level 0 packages(2.7\%).
When looking at the more prolific users, the fraction of issues created for level 1 packages increases further, and the fraction of issues created for level X packages gets reduced, while the other two remain almost similar. This indicates they are more likely to create issues for their direct dependencies and less likely to create issues for packages none of their public repositories depend on, while their likelihood of creating issues for level 0 and level 2+ packages remain similar to the likelihood for all users.

We also decided to look at the conditional distributions of issues, that are created by users who have created at least one issue to a particular level in their respective supply chains. We noticed that the fraction of issues created for level 1 packages is significantly increased when we focus only on the users who have created at least one issue for a level 0 or level 1 package. Looking at the users who created at least one cross-project issue, the fraction of issues created for level 1 packages is still increased, but by a lesser amount, while the fraction is reduced when we focus on users who created at least one issue for a level X package. This indicates the users who create issues for a level X package are likely different from the rest, which we investigate further while answering RQ2.

While looking at the distribution of pull requests (Table~\ref{t:pr}), we see a trend very similar to the one we saw for the issues, with the fraction of PRs created for level 1 being even larger under all condition, and the fraction being smaller for level X packages. The fraction under the different conditions also follow a trend similar to what saw for issues. 

In summary, looking at the distribution of issues, we notice that most of the issues are created for the users' direct dependency packages, but a number of issues are also created for packages on which none of the users' public repositories depend on even transitively, which wasn't something we expected. As for pull requests, we see more of them being created for level 1 packages, but again, a number of PRs are being created for the level X packages. When looking at the more prolific users, we see even more issues and PRs being created for level 1 packages, and less issues/ PRs being created for level X packages, but the fraction of issues/PRs being created for level 0 or level 2+ packages don't change by much. Also, we observed very few cross-project issues, and even fewer cross-project PRs under all conditions.

\vspace{-10pt}
\subsection{{\normalsize Can we identify different groups among the users based on their participation patterns? (RQ2) \texorpdfstring{\newline}{} Does the distribution change when we look at the more prolific users? (RQ3) }}

\begin{table}
\caption{No. of members and Probabilities of creating issues at different levels for the cluster centers for Cases I and III}
\label{t:c1}
\vspace{-10pt}
\resizebox{\linewidth}{!}{%
\begin{tabular}{ p{2.5cm}| p{1.25 cm}|p{1.25 cm}|p{1.25 cm}|| p{1.25 cm}|p{1.25 cm}|p{1.25 cm}}
\toprule
 & \multicolumn{3}{c||}{Case I} & \multicolumn{3}{c}{Case III} \\ \hline
 & Cluster 1 & Cluster 2 & Cluster 3 & Cluster 1 & Cluster 2 & Cluster 3 \\ \midrule
No. of members & 78047 (29\%) & 8520 (3\%) & 185575 (68\%) & 5612 (29\%) & 8932 (46\%) & 4832 (25\%) \\ \midrule
Probability of creating issue in level 0 & 0.002 & 0.01 & 0.001 & 0.02 & 0.01 & 0.004 \\ \hline
Probability of creating issue in level 1 & \textbf{0.952} & 0.03 & 0.007 & 0.53 & \textbf{0.89} & 0.050 \\ \hline
Probability of creating issue in level 2+ & 0.006 & \textbf{0.92} & 0.001 & 0.13 & 0.02 & 0.012 \\ \hline
Probability of creating issue in level X & 0.040 & 0.04 & \textbf{0.991} & 0.32 & 0.08 & \textbf{0.934} \\ \bottomrule
\end{tabular}%
}
\vspace{-10pt}
\end{table}

\begin{table}
\caption{No. of members and Probabilities of creating PRs at different levels for the cluster centers for Cases II and IV}
\label{t:c2}
\vspace{-10pt}
\resizebox{\linewidth}{!}{%
\begin{tabular}{p{2.5cm}|r|r||r|r}
\toprule
& \multicolumn{2}{c||}{Case II} & \multicolumn{2}{c}{Case IV} \\ \hline
 & Cluster 1 & Cluster 2 & Cluster 1 & Cluster 2 \\ \midrule
No. of members & 58826 (67\%) & 28827 (33\%) & 12842 (80\%) & 3127 (20\%) \\ \midrule
Probability of creating PR in level 0 & 0.007 & 0.01 & 0.01 & 0.04 \\ \hline
Probability of creating PR  in level 1 & \textbf{0.974} & 0.02 & \textbf{0.95} & 0.14 \\ \hline
Probability of creating PR in level 2+ & 0.007 & 0.02 & 0.01 & 0.04 \\ \hline
Probability of creating PR in level X & 0.012 & \textbf{0.95} & 0.03 & \textbf{0.78} \\ \bottomrule
\end{tabular}%
}
\vspace{-15pt}
\end{table}

\begin{figure}[!t]
\centering
\vspace{-10pt}
\includegraphics[width=0.7\linewidth]{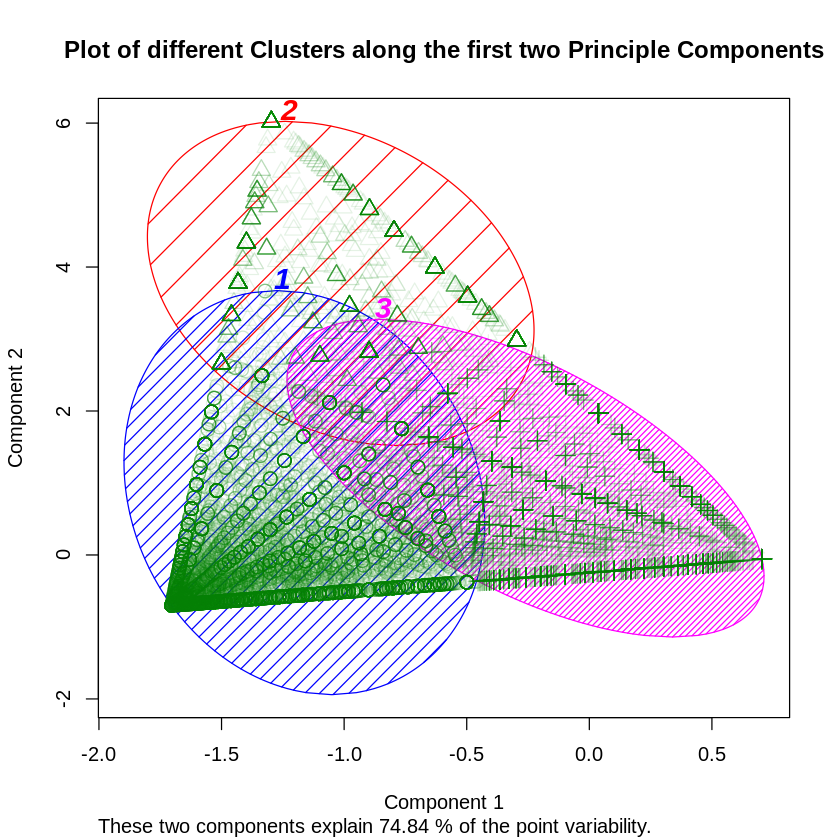}%
\caption{Visual Representation of the 3 clusters for Case I}
\label{fig:clus}
\vspace{-20pt}
\end{figure}

We discussed the analysis method used to answer this research question in Section~\ref{ss:an}. We ran the fuzzy c-means clustering algorithm 4 times, once with the marginal probabilities of all users creating an issue (Case I), and once with the marginal probabilities of users, who have created at least one PR, creating a PR (Case II) at different levels of their respective supply chains. Then we repeated the same with the users who have created 10 or more issues (Cases III and IV). For the sake of brevity we only show the visual representation of the clusters created for all users' probabilities of creating issues (Case I). The others \textit{will be available in our GitHub repository later, along with our code and other results}.

Looking at the result of clustering, we noticed 3 different clusters for Cases I and III, however, for Cases II and IV, we found two major clusters. We show a visual representation of the clusters created for Case I in Figure~\ref{fig:clus}, where the data points (in \textit{green}) are plotted along the first two principle components, and the three clusters are shown as the three shaded regions. Since the first two components explain around 75\% of the data, we assume this is a fairly accurate representation. 

We show the number and percentage of data points in each cluster, along with the cluster centers for Cases I and III in Table~\ref{t:c1}, and for Cases II and IV in Table~\ref{t:c2}. Since we used the probabilities of users creating issues and PRs as our data source, the cluster centers indicate at which level of their respective supply chains the users in that cluster are more likely to contribute issues and PRs to.

Looking at Table~\ref{t:c1}, we notice that for Case I, more than 2/3rds of all the users (cluster 3) belong to the group who are very likely to create issues for  packages in level X, around 29\% of the users (cluster 1) are avid contributors to their direct dependencies (level 1), and a small group of users (3\%, cluster 2) also exists who contribute heavily to their transitive dependencies (level 2+), i.e. they are very likely to create cross-project issues. For the more prolific users (Case III), we see a slightly different picture. Although we again see a group of users who contribute heavily to their level X projects (cluster 3), the percentage of the users is reduced to only 25\%, while the population of users who contribute heavily to level 1 projects (cluster 2) now consist of around half (46\%) of the population. Once again, we see a group of users (around 29\%) who are much more likely than the overall population average to contribute cross-project issues (cluster 1), but these users also contribute a lot of level 1 issues, and some level X issues as well. 

From Table~\ref{t:c2}, we notice that 2/3rds of the users (cluster 1, Case II)) who have created at least one pull request are very likely to create them for their direct dependencies, while the rest (cluster 2) are more likely to create issues for their level X dependency packages. Looking into the more prolific users (Case IV), we notice that the percentage of users who are likely to create PRs to level 1 packages (cluster 1) is increased to 80\%, while the other 20\% (cluster 2) are more likely to create PRs to level X packages, but they also create a number of PRs for level 1 packages, and are more likely to create PRs for level 0 and 2+ packages.

We examined the amount of activities of different users belonging to different clusters and found that the users who commit more to their direct dependencies are more active, creating more issues and PRs, and committing to more repositories, while the users more likely to commit to level X packages show very little activity and many of them have company affiliations. The users who are likely to create cross-project issues tend to have a large number of transitive dependencies, and create very few PRs. All of these differences were significant, which was verified using the Kolmogorov-Smirnov test.

In summary, we see three different groups of users based on which level of their respective supply chains they create issues for. While a large number of users are likely to create issues for level X packages, a group consisting of a good number of users are more likely to create issues for level 1 packages, and a small group of users also exists who are likely to create cross-project issues. In terms of creating PRs, we see two major group of users: 2/3rds of the users are more likely to create PRs to level 1 packages, while the rest are more likely to create PRs for level X packages. Looking into the more prolific users, we again see three groups of users based on their issue creation patterns, but the percentage of users who create issues for level X packages is reduced, and the fraction of users who create issues for level 1 packages is increased. As for the users who created at least one PR and 10 or more issues, the fraction of users belonging to the group who are very likely to create PRs to level 1 increase even further, while the rest of the users form a group who are more likely to contribute PRs to level X packages.

\vspace{-10pt}
\subsection{ {\normalsize Using participation patterns of users to identify their company affiliation (RQ4)}}

\begin{figure}[!t]
\centering
\vspace{-10pt}
\includegraphics[width=0.6\linewidth]{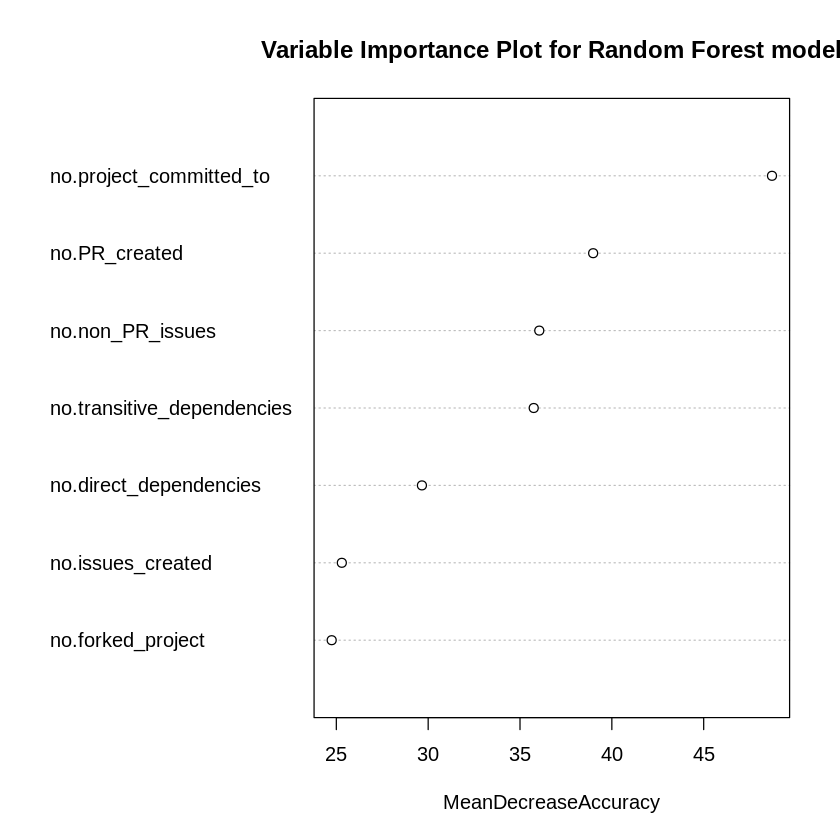}%
\caption{Variable Importance plot of the Random Forest model}
\label{fig:rf}
\vspace{-20pt}
\end{figure}

To answer this question, we used Random Forest modeling technique, as mentioned in Section~\ref{ss:an}. Our dataset had the predictors at listed in Table~\ref{t:var}. We dropped the predictor ``User.login'', and were left with 30 predictors and our response variable was the binary variable representing if the user had a company affiliation. To obtain the optimal number of predictors we used the ``rfcv'' function from the \textit{randomForest} R package, which shows the cross-validated prediction performance of models with sequentially reduced number of predictors (ranked by variable importance) via a nested cross-validation procedure.
Looking at the output of this function, we decided to use 7 predictors for our final model. 

First, we created a Random Forest model with all the predictors, and selected the top 7 predictors by looking at the variable importance plot. 
To calculate the performance of the model, we decided to use 70\% of the data, selected randomly, as our training set, and the other 30\% as our test set.
Then, to optimize our model, we decided to tune the model parameters, viz. ``mtry'', the number of variables randomly sampled as candidates at each split, and ``ntree'', the number of trees to grow. 
We used the ``train'' function from the \textit{caret} package in R for performing a grid search on the training data to find the optimal values of the two parameters that gives the highest Accuracy, using 10 fold cross-validation.
The optimal value of ``mtry'' was found to be 2, and ``ntree'' of 500 gave the best performance.

Using the optimal values of the parameters ``mtry'' and ``ntree'', we fitted the Random Forest model on the training data, and tested the performance of the model against the test data. Our model had a sensitivity of 0.62, and it performed relatively worse in terms of specificity (0.47), i.e. it did relatively better in terms of not classifying users without a company affiliation as users with a company affiliation, but a number of users with a company affiliation were wrongly predicted as users without a company affiliation. The value of AUC under the ROC curve was 0.68, and the overall accuracy of our model was 0.70, with a 95\% confidence interval between 0.69 and 0.75. 

The variable importance plot for our final model is shown in Figure~\ref{fig:rf}. The 7 predictors we selected for our final model were (in the same order of importance they appear in Figure~\ref{fig:rf}): total no. of Git repositories a user committed to, no. of pull requests created by the user, no. of issues created by the user that are not pull requests, total number of transitive dependencies of all of the user's public repositories, total number of direct dependencies of all of the user's public repositories, total number of issues created by a user, total no. of repositories of the user that are forks of another repository. So, we see that to which layer a user creates an issue or a pull request isn't really important in predicting their company affiliation, but the total activity, the number of projects they committed to, the number of issues, PRs, and non pull request issues they create, and the number of packages the user's public repositories depend on directly and transitively are important in predicting their company affiliation. 

To observe how the values of these predictors are different between users with and without a company affiliation, we conducted the one-sided Kolmogorov-Smirnov test to test if the distribution is stochastically larger for one of the groups, for these variables. We found that for users with a company affiliation, the distributions of all of the predictor variables are stochastically larger, i.e. they create more issues, more PRs, as well as more non PR issues, and they also commit to more projects, have more dependencies, and more forked projects. Overall, we can say that they have a larger footprint on the NPM ecosystem. 

\vspace{-10pt}
\section{Discussion}\label{s:disc}

In this section, we discuss the answers we obtained for our research question, and the implications of our findings. The important findings of our study include: (1) The distribution patterns of issues and PRs for the NPM ecosystem, which highlight that there are very few cross-project issues and PRs. (2) The presence of distinct user groups, who differ significantly in their participation patterns and amount of activity, and the existence of a large number of users who contribute to packages in level X. (We expected some users like this, since some of their activity may not be public, but we didn't expect so many users would be part of this group.) (3) The shift in participation patterns for the more prolific users, and (4) The possibility of predicting the users' company affiliation by their participation patterns.

Our RQ1 was focused on the distribution of the total number of issues created, and our RQ2 investigated the existence of different groups of users based on the distribution of probabilities of them creating issues at different levels of their respective supply chains. We observed that in terms of creating issues, only 29\% belonged to the group who are more likely to create issues for their direct dependencies, but they create around 53\% of the total issues. An opposite picture was observed for users who create issues for level X packages, where 68\% of the total users are likely to create issues for those packages, but they create around 40\% of the issues. This indicates the users who create issues for their direct dependencies are more active. This assumption is further validated when we look at the more prolific users, which shows that more of the prolific users are likely to create issues for their direct dependencies. We observe a similar pattern when we focus on the distribution of PRs and the users who create PRs. However, in this case, we have more users in the group of those more likely to create PRs for level 1 packages. Users creating more issues and PRs for their direct dependencies isn't surprising, since they might face more issues from them and feel more obliged to fix the issues in those packages. However, the overall trend observed while answering RQ1 and RQ2 led to the following possible implications:
(1) The users who create demand mostly from their direct dependencies are different in nature from those who create demand (issues) from packages outside their supply chain, given they belong to different clusters, and they also differ in their amount of activity. A study looking into the differences between the two groups, their nature, motivation, and reasons for their distinct contribution patterns might give new insights into the NPM ecosystem.
(2) We can assume the users who submit PRs are, on an average, more technically proficient than the rest, at least in the given domain. Given the prevalence of low quality issues~\cite{xie2013impact}, it might be helpful to predict the quality of an issue or a pull request using the contribution pattern of the user who submitted it. 

We observed very few cross-project (level 2+) issues, and even fewer cross-project pull requests. We hypothesize that the reason behind this is a mixture of two factors, (1) the users may not be aware which package is causing some issue they are facing or they do not know how to go about fixing the issue, and (2) they might feel it is not their responsibility to report or fix those issues. A similar situation was reported in~\cite{valiev2018ecosystem}, which studied the PyPi ecosystem, where a developer said that their experience in trying to fix a bug just two levels upstream was ``Extremely Painful'', due to their unfamiliarity with the issue reporting system and resolving process, and not being able to convey their problem clearly to the developers in charge. We suspect a similar situation could be true for the NPM ecosystem as well.
So, if the reason behind the users not reporting and fixing cross-project issues is more due to the lack of transparency, then this calls for the need of tools and practices that would increase the visibility for the developers beyond the direct dependencies of their code and that would help determine how the packages far in the supply chain might be affecting some issues that  they discover when running their code.
However, we did observe a small group of users who are more likely to create cross-project issues, both for all the users and the more prolific users, but such a group was not observed when investigating pull requests. Investigating those users might be helpful in formulating a way to increase visibility and streamline the cross-project issue reporting process.

We observed that users with a company affiliation, overall, are more active than the rest, i.e. they contribute to as well as demand more effort from the projects, which might mean that the involvement of different companies is a major driving force behind the growth of the NPM ecosystem. So, if an NPM package gets supported/used by a company, it might be beneficial for the growth of that package, and of the NPM ecosystem overall. Does it indicate the FLOSS community is shifting from its initial structure of software by and for the users~\cite{von2001learning}? That is a much larger question that needs further study to answer, but our result indicates that companies might have a larger impact on the NPM ecosystem. Using a model similar to ours for identifying the commercial affiliation of the users, and identifying the differences in their contribution patterns might be useful for answering that bigger question.

\vspace{-10pt}
\section{Limitations}\label{s:limit}

There are a few limitations to our study that we would like to highlight here. First of all, we only considered the Git repositories with a \texttt{package.json} file as JavaScript projects, which is not always true. Also, we extracted the dependency information by looking at the \texttt{package.json} and \texttt{lerna.json} files, however, looking directly into the source code might have given a much more accurate picture of dependencies. As for dependencies, the dependency map we constructed is for runtime dependency only, i.e. we did not consider the \textit{devDependencies} or any other type of dependencies . 

We have assumed in this study that issues create a demand of effort to fix it, and pull-requests can be regarded as contribution of effort by the developers who use a package. While this might be true in general, there is definitely the possibility that the maintainers of a project end up spending a lot of effort fixing some pull-request of poor quality, and, on the other hand, creating a good quality issue report also takes effort from the part of an issue reporter, and the maintainers might have to spend little effort fixing an issue of good quality. However, we believe that our assumption holds true for majority of the cases.

We only looked at the public repositories of the users, for obvious reasons. So, it could be possible that, based on the activity of a user in their private repositories or other projects not shared publicly in Git, some of the packages that we classified as level 2+ for a user could actually be level 1 for them, or some package in level X could actually belong to level 0, 1, or 2+ for that user. 

We looked at only 4433 NPM packages, which is less than 0.5\% of the total packages in NPM ecosystem, however, given that a huge number of packages are almost never used, we believe this small subset of packages experience bulk of the activity in the ecosystem. 

As mentioned before, we extracted the company affiliation information for the users from the information they provided on GitHub. We did not attempt to validate this information from any other source, which leaves the room for some error in classification. However, we believe that more professional developers are likely to provide accurate information about themselves. Another related situation could be that some users actually affiliated to a company never bothered to fill out that information about themselves, leading to a misclassification.

While studying the issues, we did not differentiate between the type of issue, if it is open or closed, and for the pull-requests, if it was merged or not, nor have we checked if the company a user is associated with is one that is centered around OSS development, or a more traditional company.

Our study selected the users based on the criteria that must have created at least one issue, which makes all of our findings are conditional on that selection criteria, and the results may not apply for the entire population of users.

The result we obtained in this paper might not generalize to all types of software ecosystems, since NPM is heavily used by different companies around the world, while many other types of software are not as heavily used. 

\vspace{-10pt}
\section{Conclusion}\label{s:conclusion}

We have separated what is typically considered to be a contribution of effort into a part that likely demands more effort from projects (issue fixes) and a part that is likely to provide more value (patches) and investigated where in the supply chain these occur and if there are distinct participation patterns. Initial findings suggest the lack of visibility and highlights groups of participants that contribute in radically different ways. Future studies are needed to determine how to increase the visibility and learn from distinct participation patterns and how these findings apply in other ecosystems.

\bibliographystyle{ACM-Reference-Format}
\bibliography{sigproc} 

\end{document}